\documentstyle[12pt,psfig]{article} 
 1                                        
\begin{document}                                                              
\begin{center}                                                                
{ \large\bf Three-body mechanism of $\eta$ production\\}
\vskip 1cm  
{K. P. Khemchandani$^1$, N. G. Kelkar$^2$ and B. K. Jain$^3$\\} 
\vskip 0.3cm  
$^1${\it Nuclear Physics Division, Bhabha Atomic Research Centre, Mumbai 400085,
India}\\
$^2${\it Departamento de Fisica, Universidad de los Andes,
Cra. 1E, No. 18A-10, Santafe de Bogota, Colombia}\\
$^3${\it Physics Department, University of Mumbai, Mumbai - 400098, India, \\
Saha Institue of  Nuclear Physics, 1/AF, Bidhan Nagar, Kolkatta, India}\\
\end{center}
\vskip .5cm
\begin{abstract}
The usefulness of the three-body mechanism of $\eta$-meson production 
is tested by extending to high energies a recently developed 
two step model of the 
$p d \rightarrow ^3$He $\eta$ reaction by us , which was successful in 
reproducing the energy dependence of the data near threshold. The 
$\eta$ production in this model, proceeds in two steps via the 
$p p \rightarrow \pi d$ and $\pi N \rightarrow \eta p$ reactions
with the intermediate particles being off-shell.  
The final state interaction(FSI) is incorporated through a $T$-matrix for 
$\eta^3$He elastic scattering constructed using few body equations. 
In a comparison of the calculated angular distributions with some recently 
available data, we discover the limitations of the model at
high energies. 
\end{abstract}
\newpage
\section{Introduction}
The importance of the three-body mechanism (3-bm) in meson production was
first shown in \cite{LL1,LL2} by Laget and Lecolley. Whereas in the pion
producing reactions $p d \rightarrow ^3$H $\pi^+$ and $p d \rightarrow
^3$He $\pi^0$, this mechanism accounted for the discrepancies left out 
at backward angles by
the one- and two-body mechanisms, in the case of the $p d 
\rightarrow ^3$He $\eta$ reaction, it was found to play the dominant role 
in reproducing the Saturne data at threshold \cite{berger} and 
that at high energies \cite{berthet} for $\theta_{\eta} = 180^o$. 
The one- and two-body mechanisms in the 
$p d \rightarrow ^3$He $\eta$ reaction, were found to underestimate the 
experimental cross sections by two orders of magnitude. Motivated by these
works, more recently, two step models \cite{wilkin,kondra,we3} which essentially
involved the 3-bm, were constructed to study the $p d \rightarrow ^3$He $\eta$ 
data near threshold \cite{berger,mayer}. 
The $\eta$-meson in these models is produced in two
steps via the $p p \rightarrow \pi d$ and $\pi N \rightarrow \eta N$ 
reactions, thus involving the participation and sharing of momentum by 
three nucleons which is typical of a 3-bm (see Fig. 1). 
The calculation in \cite{we3} is the most involved one available on 
the $p d \rightarrow ^3$He $\eta$ reaction in literature.
In it, (i) the 
final state $\eta^3$He interaction which is crucial for the energy 
dependence of this reaction is incorporated through a T-matrix constructed
using few-body equations and (ii) the fermi motion and off-shell nature
of the intermediate particles in the 3-bm is taken into account properly.
Some other theoretical investigations of the $p d \rightarrow ^3$He $\eta$ 
reaction 
at threshold and high energies can also be found in \cite{others,stenmark}. 
The work in \cite{stenmark} needs a special comment to which we shall 
return later. 
\begin{figure}[h]
\centerline{\vbox{
\psfig{file=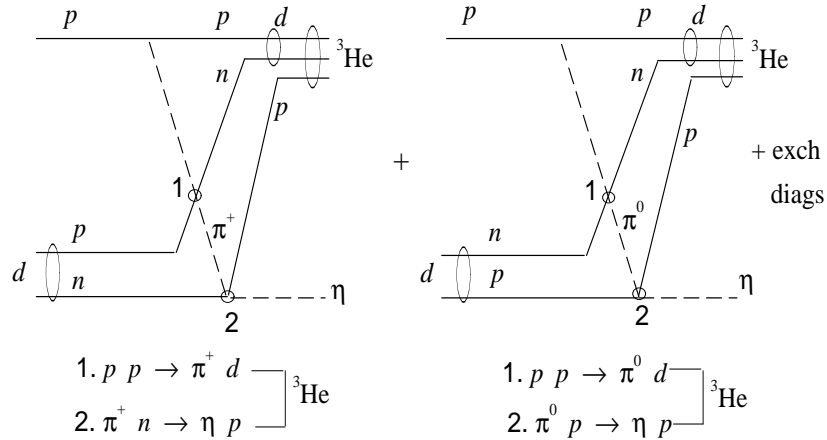,height=6cm,width=11cm}}}
\caption{The diagram corresponding to the three-body mechanism of 
$\eta$-meson production in the $p d \rightarrow ^3$He $\eta$ reaction.}
\end{figure}

The explanation for the dominance of the 3-bm in $\eta$ production is
straightforward. Due to the large mass of the $\eta$-meson ($\sim$ 547 MeV), 
the momentum transfer involved in this reaction is as large as 900 MeV/c
near threshold. The 3-bm allows this large momentum to be shared 
among three nucleons, thus making it easily digestible by the nucleus than 
in the one- and two-body 
mechanisms where the nuclear form factor (evaluated at large momenta) is
rather small. Since the two step model alone 
could reproduce the strong energy dependence of the 
threshold data \cite{we3} and given the fact that the momentum transfer in the 
$p d \rightarrow ^3$He $\eta$ reaction continues to be large ($\sim$ 650 MeV/c) 
even at 1 GeV above threshold, one would speculate 
that the same model would also be successful in reproducing the recently
reported angular distributions \cite{bilger} at high energies. 
An extension
of the model in \cite{we3} to compare with the high energy data of 
\cite{bilger}, therefore, seems timely.

However, before proceeding further, let us recapitulate the following 
observations from the different 3-bm calculations carried out in the
literature at energies beyond threshold: 
(a) the success of the 3-bm reported in \cite{LL2} for the 
the $p d \rightarrow ^3$He $\eta$ reaction over a range of proton beam 
energies up to 2.5 GeV, was confined to one particular 
angle, namely, $\theta_{\eta}=180^0$, (b) in case of pion production at
high energies, the 3-bm was mostly responsible in removing the discrepancies
between theory \cite{LL1} and data at backward pion angles,  
(c) the theoretical estimates \cite{kondra,komar} 
of angular distributions (at threshold and at high energies) of the 
$p d \rightarrow ^3_{\Lambda}$H $K^+$ reactions made
within a two step model, seem to indicate backward peaked 
cross sections and (d) the cross sections calculated in \cite{we3} for the 
$p d \rightarrow ^3$He $\eta$ reaction  
within the two step model were nearly isotropic near threshold; however, 
there was a hint of a backward peak as one moved about 10 MeV away from 
threshold.

That is, in all, the existing 3-bm calculations point toward a backward
angle dominance of the cross sections in this model. In contrast to this,  
the measured angular distributions of the $p d \rightarrow ^3$He $\eta$ 
reaction from 
about 90-200 MeV above threshold are all forward peaked \cite{bilger,cosy}. 
Thus, a priori, one would doubt the possibility of  
the two-step model being able to
reproduce the high energy data. Indeed in \cite{zloman}, in an attempt
to reproduce the differential cross section data at beam energies of
930, 965, 1037 and 1100 MeV, using a Monte-Carlo description of the 
$p d \rightarrow ^3$He $\eta$ reaction within a two-step model, the authors
find backward peaked cross sections which are in 
complete disagreement with data, but are however, in line with the above theoretical
findings. 
A very recent work \cite{stenmark}, which  
reports the success of the two step model in reproducing the data on 
angular distributions mentioned above, is then isolated in its findings. 
The results in \cite{stenmark} have been obtained 
using a factorization approximation, which neglects the off-shell nature of 
the intermediate pion propagator as well as the t-matrices for the 
$\pi N \rightarrow \eta N$ and $p p \rightarrow \pi^+ d$ reactions. 
It restricts the pion in an ad hoc way to go at $0^0$. Besides, though
not mentioned clearly, it seems that the final state interaction has
also been incorporated within a factorization approximation as in 
\cite{wilkin}. 
The model with such approximations, does not truly represent the three-body 
Feynman diagram of Fig. 1 (where the intermediate particles are off-shell), 
hence the results of such specious models should be treated with 
caution. 
To obtain a proper estimate of the 
contribution of the 3-bm at higher energies, we extend our earlier 
calculation \cite{we3} 
to evaluate the differential cross sections 
at 930, 965, 980, 1037 and 1100 MeV beam energy for which data are available. 
These calculations include the off-shell and final state interaction effects
elaborately and do not restrict the pion to any specific angle.
Unlike \cite{stenmark} and as expected, partly from our earlier work 
\cite{we3} and partly from 
calculations of other reactions with 3-bm in literature 
(mentioned in (c) above), 
we get backward peaked angular distributions which are at variance with 
the corresponding forward peaked data.  
The cause for this discrepancy may lie in the interaction among
the particles in the intermediate state or the increasing importance
of some other mechanisms which need to be investigated. We therefore 
conclude that the 3-bm calculations of the 
$p d \rightarrow ^3${\rm He}$\,\eta$ reaction at energies away from the
threshold do not reproduce the observed forward peaking of the
$\eta$ angular distribution seen in the Uppsala data. This is in
contradiction with the observations reported in \cite{stenmark}.

In the next section we describe the inputs of the model only briefly and 
move on to the discussion of the results. 
The details of the model can be found in \cite{we3}.  

\section{Two step model of the $p d \rightarrow ^3${\rm He}$\,\eta$ reaction}

In the two step model of $\eta$ production, the incident proton 
interacts with one of the nucleons in the deuteron ($d$) to produce a 
deuteron ($d^{\prime}$) and pion, which in turn interacts with the other
nucleon in $d$ to produce an $\eta$-meson and a proton. This proton 
and the $d^{\prime}$ combine to form the $^3$He nucleus. The transition
matrix for the $p d \rightarrow ^3$He $\eta$ reaction which includes 
the final state $\eta^3$He interaction is given as,
\begin{eqnarray}\label{tmat2}
&&T =\, <\,\vec{k_\eta}\, ; \, m_3\,|\, T_{p d \rightarrow \,^3He
\,\eta}\,|
\,\vec{k_p}\, ; \, m_1 \, m_2\,> + \\ \nonumber
&&\sum_{m_3^\prime} \int { d\vec {q} \over (2\pi)^3} {<\, \vec{k_\eta}\,
; \, m_3\,|
\, T_{\eta\, ^3He}\, |\,\, \vec{q}\, ; \, m_3^\prime\,> \over E(k_\eta)\,
- \,E(q)\,
+\, i\epsilon}\,
<\vec{q}\, ; \, m_3^\prime\,| T_{p d \rightarrow \,^3He \,\eta}\,|
\,\vec{k_p}\, ; \, m_1 \, m_2>. 
\end{eqnarray}
The matrix elements $< |T_{p \,d \rightarrow \,^3He\, \eta}| >$ in the
above equation correspond to the Born amplitude for the
$p\, d \rightarrow \, ^3$He$\, \eta$ reaction. 
$ \vec {k_p}$ and $ \vec {k_\eta}$ are the asymptotic momenta
of the particles in the initial and final states and $m_1$, $m_2$ and 
$m_3$ are the spin projections of the proton, deuteron and $^3$He 
respectively. The $T$-matrix for $\eta^3$He 
elastic scattering, $T_{\eta\,^3He}$, is evaluated using four-body equations
for the $\eta$(3N) system. The Born amplitude is given within the 
two step model as,
\begin{eqnarray}\label{ampli}
< |T_{pd \rightarrow ^3He\,\eta}| >=i \int {d\vec{p_1}\over (2\pi)^3}
{d\vec{p_2}\over (2\pi)^3} \sum_{int\,m's} <p n \,|\,d>
\,<\pi\,d |T_{pp\, \rightarrow\, \pi\, d}| p\,p>
\\ \nonumber
 \times{1\over (k_\pi^2-m_\pi^2+i\epsilon)}
\, <\eta\,p \,|\,T_{\pi N \rightarrow \eta p}\,| \pi\,N>
\,\,<\,^3He\,|\,p\,d>\,,
\end{eqnarray}
where the sum runs over the spin projections of the intermediate
off-shell particles. The four momentum of the intermediate pion, $k_{\pi}$, 
which could either be a $\pi^+$ or $\pi^0$ is given as, 
\begin{eqnarray}
\vec{k}_\pi &=& {\vec{k}_p \over 2} + {2\over 3} \vec{k}_{\eta} + \vec{p}_1 
+ \vec{p}_2 \\ \nonumber
k_{\pi}^0 &=& E_{\eta} + {1\over 3} E_{^3He} - {1\over 2}E_d\,,
\end{eqnarray}
where $\vec{p}_1$ and $\vec{p}_2$ are the fermi momenta of the intermediate
nucleons.    
Each of the individual matrix elements in the above equation
is expressed in terms of partial wave expansions. The matrix elements
for the $p p \rightarrow \pi d$ reaction, parametrized in terms of
the available experimental data are taken from Ref. \cite{arndt}. For
the $\pi N \rightarrow \eta p$ reaction, we use the
single resonance, coupled
channel model of Ref. \cite{batinic}. This model treats 
the $\pi N$, $\eta N$ and
$\pi \pi N$ as coupled channels. 
We use the $T$-matrix in this model with the inclusion of the 
$S_{11}$, $P_{11}$ and $D_{13}$ resonances.  
Since the $\eta-P_{11}$ coupling used in Ref. 
\cite{batinic}, as noticed in \cite{stenmark}, 
does not reproduce the data on $\pi \, N \, \rightarrow \, \eta 
 \, N$ reaction, we use it as given in \cite{stenmark}.  
The matrix elements $<p n|d>$ and $<^3$He$|p d>$ consist of 
the deuteron and helium wave functions in momentum space which depend on 
the momenta $\vec{p}_1$ and $\vec{p}_2$ respectively. The details of the 
wave functions and the $\eta^3$He $T$-matrix are given in \cite{we3}. 
In the calculation of the $\eta^3$He $T$-matrix, we use the coupled channel 
elementary $t$-matrix, 
$t_{\eta N \rightarrow \eta N}$ as in \cite{fix}.  
This $t$-matrix was the one
which gave the best agreement in \cite{we3} with the threshold data.

\section{Results and Discussion}
Recent measurements of the $p \, d \, \rightarrow \, ^3$He$\, \eta$ 
reaction, from about 22 MeV to 115 MeV above threshold \cite{bilger, cosy} 
show that, as we go away from the threshold, the angular distribution
starts developing anisotropy, which increases with energy.   
The GEM data \cite{cosy} show a strong forward peak and the 
data from the
Wasa-Promice collaboration \cite{bilger} are found to be maximal for 
cos $\theta_\eta \, \approx $+0.5. 
At threshold, the angular distribution is isotropic. 
\begin{figure}[ht]
\centerline{\vbox{
\psfig{file=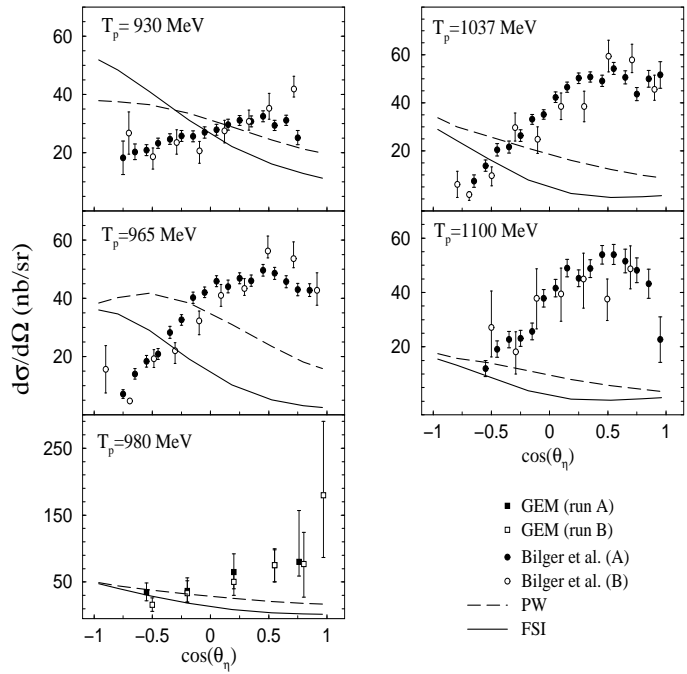,height=9cm,width=9cm}}}
\caption{ The angular distribution of the $p \, d \, \rightarrow \, ^3$He$\, 
\eta$ reaction. The data are from Ref. \cite{bilger,cosy}. The data points 
(A) and (B) of \cite{bilger}, represent respectively, the data taken with 
only the $^3$He detected and with $^3$He detected in coincidence with two 
photons. The dashed lines (solid lines) are calculations without (with) 
inclusion of the final state interaction (FSI).}  
\end{figure}

Fig. 2 shows our calculated differential cross 
sections,  
using the two step model described in the previous section, 
along with the experimental data from 
\cite{bilger, cosy}. The dashed curves represent the plane wave
calculations, while the solid lines correspond to the calculations including 
the $\eta^3$He final state interaction (FSI). 
We find that the observed trend in the anisotropy of the data is not at
all reflected in the calculated results. Unlike the data, the latter peak
at backward angles. This in contrast to the findings in \cite{stenmark}. 
\begin{figure}
\centerline{\vbox{
\psfig{file=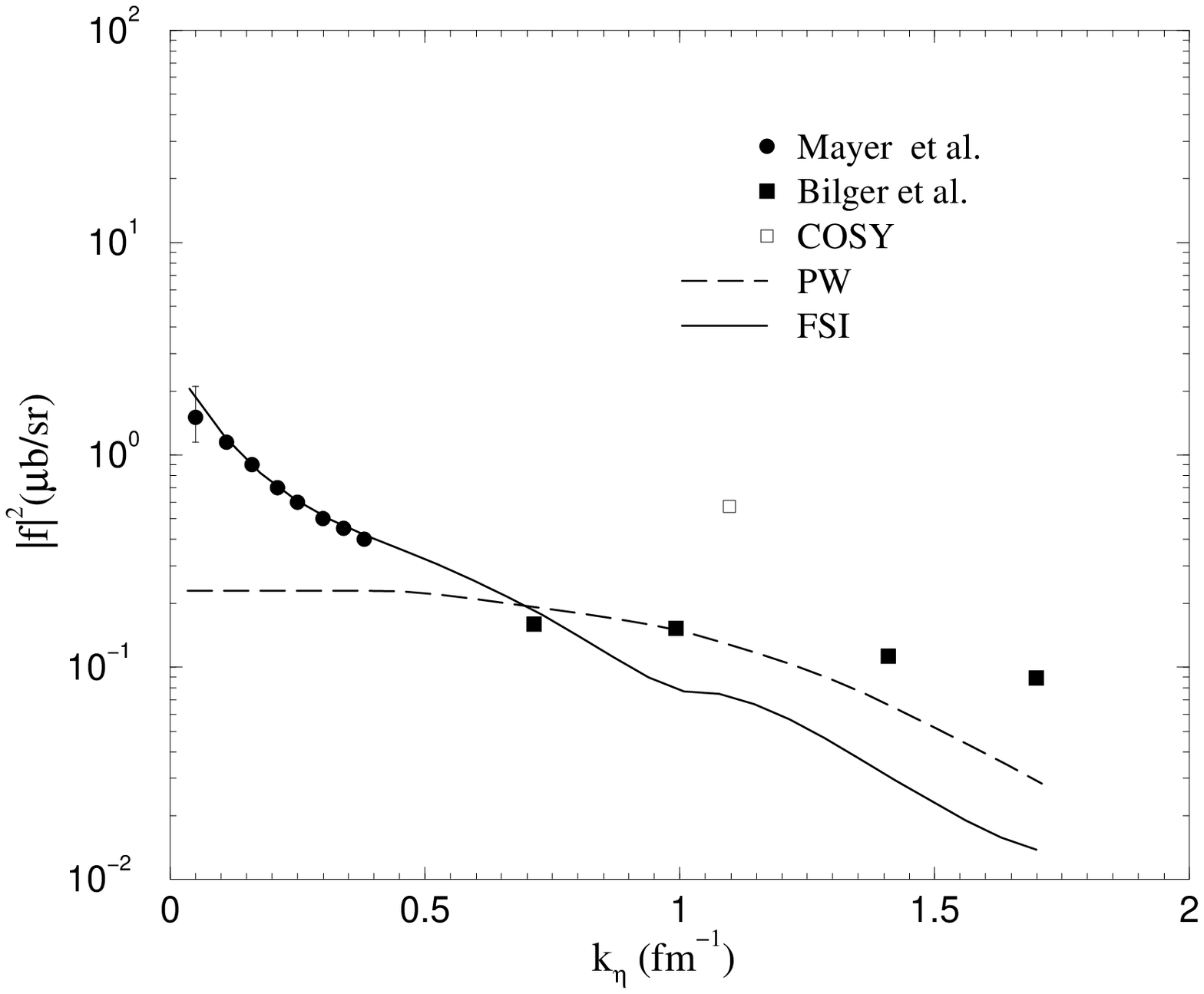,height=6cm,width=7cm}}}
\caption{The square of $p \, d \, \rightarrow \, ^3$He$\, \eta$ amplitude as  
defined in eq. (\ref{fsq2}). The dashed (solid) lines correspond
to the calculations without (with) FSI. The data are from refs 
\cite{mayer,bilger,cosy}.}  
\end{figure}

Fig. 3 shows the data and calculation of the spin averaged amplitude
which is defined as,
\begin{equation}\label{fsq2}
\mid f \mid^2 \, = \, {k_p \over k_\eta}\,{\sigma_{tot} \over 4\pi}\, ,
\end{equation}
where $k_p$ and $k_\eta$ are the proton
and  $\eta$ momentum in the center of mass system. 
$\sigma_{tot}$ is the angle integrated cross section. In the 
experiment it is obtained by making a polynomial fit to the data at the
available angles and then integrating the parameterized distribution. 
We observe two things: (i) the two step model results are in excellent
accord with the data near threshold. The model however underestimates
the data at higher energies. (ii) Comparison of the calculated results
with and without FSI show that, while near threshold the FSI increases
the cross section, at higher energies its effect is to decrease the
cross sections. This, as discussed in our earlier work \cite{we3}, 
is the reflection of the decreasing importance of the off-shell 
$\eta - ^3$He scattering in the final state at high energies.

\begin{figure}
\centerline{\vbox{
\psfig{file=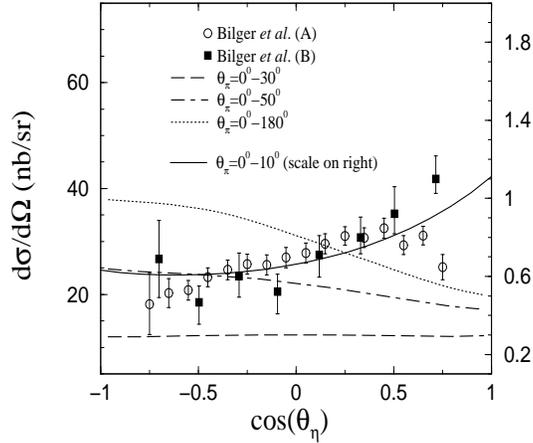,height=6cm,width=7cm}}}
\caption{The angular distribution for the $p d \rightarrow ^3$He $\eta$ 
reaction at 930 MeV beam energy. The solid line shows the cross section
with the intermediate pion angle restricted from $0$ to $10^0$ with its 
scale on the right hand side. 
The long dashed, dot-dashed and dotted curves show the cross sections obtained 
after restricting the pion angle from $0$ to $30^0$, $0$ to $50^0$ and
allowing the full range of $0$ to $180^0$ respectively (with the
scale given on the left).} 
\end{figure}
To see the implication of the restriction on the direction of the
intermediate pion, we show in Fig. 4, our calculated results at one
beam energy, for pions restricted up to various emission angles.
As found in \cite{stenmark}, the forward pion emission indeed
skews the differential cross section to forward angles. As such, this
restriction is of course ad hoc. Any legitimacy in this may be sought
in the interaction among the particles in the intermediate state. 
We also see that the cross sections fall with reduction 
in the range of allowed pion angles. The scale given on the right
side of Fig. 4 for the $0$ to $10^0$ restriction (solid line) shows
that though the shape of the experimental angular distribution is 
correctly reproduced, the magnitude is underestimated by about 
a factor of 40. This is again in contrast to the finding in 
\cite{stenmark}, where the agreement with absolute cross sections, 
with the pion angle restricted to $0^0$ is not so bad. 
The reason for this discrepancy could lie in the
neglect of the off-shell effects at the intermediate vertices in 
the reaction mechanism in \cite{stenmark}. 
Since we only wish to demonstrate in Fig. 4 the effect of restricting
pion angles, these calculations have been done without including the FSI. 

\section{Conclusions}
The role of the three-body mechanism had been demonstrated unambiguously
in our earlier work on the $p d \rightarrow ^3$He $\eta$ reaction near 
threshold \cite{we3}. Indeed, isotropic angular distributions and the 
magnitude of the total cross sections were well reproduced. Extending the
model to higher energies (without making any rough approximations 
outside the spirit of the model) leads to a large discrepancy between
model predictions and available data. We believe that any ad hoc 
manipulations of the model \cite{stenmark}, 
like e.g., restricting the pion to be on-shell
and to be produced in the forward direction ($\theta_{p\pi} = 0$),   
do not represent the three-body mechanism in the context of the Feynman 
diagram and hence are hard to justify. Had these conditions really been
the preferred ones in the three-body mechanism, it should have come out
anyway through a proper calculation as performed in the 
present work. However, this is not the case. 

The failure of the model, especially in reproducing data at forward
angles hints toward some missing component in the model.   
In the present
work, we do not take into account the interaction of the intermediate
off-shell particles which in principle could be quite important. 
Contribution of coupled channel processes such as the 
$p d \rightarrow ^3$He $\pi^0$ may also turn out to be significant.  
The possibility of other mechanisms, such as the one- and
two-body mechanisms taking over at high energies seems unlikely for
the following reasons: (i) the momentum transfer 
in the $p d \rightarrow ^3$He $\eta$ reaction, as mentioned in the
introduction, continues to be quite large even at high energies (ii)
as far as kinematics is concerned, the 
$p d \rightarrow ^3$He $\eta$ and $p d \rightarrow ^3_{\Lambda}$H $K^+$ 
reactions are similar. Hence, the results of the theoretical 
investigation of the $p d \rightarrow ^3_{\Lambda}$H $K^+$ reaction up 
to proton beam energy of 3 GeV \cite{komar}, where the authors find the 
one- and two-body mechanisms to contribute 2-3 orders of magnitude 
lesser than the three-body mechanism at all angles, taken along with
similar findings of \cite{LL2} for $p \vec{d} \rightarrow ^3$He $\eta$ at 
$\theta_{\eta}=180^0$, any expectations from
the one- and two-body diagrams for $\eta$ production in 
reproducing the high energy angular distribution data for the 
$p d \rightarrow ^3$He $\eta$ reaction seem remote.
Though the effect of the final state interaction has not been
studied with the one- and two-body mechanisms, from the present
work it seems that at high energies this will only further
reduce the cross sections (see Fig. 3 for example), making
the one and two-body mechanisms even more negligible.  

The above conclusions could probably change if the deuteron and 
the $^3$He have some hitherto unconsidered dynamics (like quarks) 
at short distances, which influence their structures. 
Our efforts in future would be to explore such avenues and 
the role of other possible diagrams such as the direct production
of the $\eta$ meson.

\end{document}